\documentclass{article}
\usepackage{geometry}
\usepackage{authblk}
\usepackage{lineno,hyperref}
\usepackage{amssymb}
\usepackage{amsmath}
\usepackage{amsthm}
\usepackage{amsfonts}
\usepackage{bm}
\usepackage{subcaption}
\usepackage{mathabx}
\usepackage{verbatim}
\usepackage{diagbox}
\usepackage{slashbox}

\usepackage{multicol}
\usepackage{color}
\usepackage{xspace}
\usepackage{pdfwidgets}
\usepackage{algorithm}
\usepackage{algpseudocode}

\usepackage{enumerate}

\makeatletter
\def\bs{\expandafter\@gobble\string\\}
\def\lb{\expandafter\@gobble\string\{}
\def\rb{\expandafter\@gobble\string\}}
\def\@pdfauthor{C.T.Do}
\def\@pdftitle{Robust multi-sensor GLMB filter: An application to multi-target tracking with bearing-only sensors}
\def\@pdfsubject{Robust multi-sensor GLMB filter: An application to multi-target tracking with bearing-only sensors}
\def\@pdfkeywords{Multi-sensor GLMB filter, robust tracking, bearing-only sensors, bootstrapping method,  Labeled Random Finite Sets.}

\providecommand{\keywords}[1]
{
  \small	
  \textbf{\textit{Keywords---}} #1
}
\usepackage{fancyhdr}
\title{Robust Multi-Sensor Generalized Labeled Multi-Bernoulli Filter}

\begin{document}

\author[1,*]{Cong-Thanh Do}
\author[1]{Tran Thien Dat Nguyen}
\author[2]{Hoa Van Nguyen}
\affil[1]{\small{School of Electrical Engineering, Computing, and Mathematical Sciences, Curtin University, Bentley, WA 6102, Australia}}
\affil[2]{\small{School of Computer Science,
The University of Adelaide, Adelaide, SA 5005, Australia}}
\affil[*]{\small{Corresponding author, \textit{thanh.docong@postgrad.curtin.edu.au; thanhdc@tnu.edu.vn}}}

\maketitle

\begin{abstract}
This paper proposes an efficient and robust algorithm to estimate target trajectories with unknown target detection profiles and clutter rates using measurements from multiple sensors. In particular, we propose to combine the multi-sensor Generalized Labeled Multi-Bernoulli (MS-GLMB) filter to estimate target trajectories and robust Cardinalized Probability Hypothesis Density (CPHD) filters to estimate the clutter rates. The target detection probability is augmented to the filtering state space for joint estimation. Experimental results show that the proposed robust filter exhibits near-optimal performance in the sense that it is comparable to the optimal MS-GLMB operating with  true clutter rate and detection probability. More importantly, it outperforms other studied filters when the detection profile and clutter rate are unknown and time-variant. This is attributed to the ability of the robust filter to learn the background parameters on-the-fly.

\end{abstract}
\keywords{Multi-sensor GLMB filter; robust tracking; bearing-only sensors; bootstrapping method;  Labeled Random Finite Sets.}

\section{Introduction}

Multi-target tracking has   captured the interest of the research community for more than fifty years with its  range of applications. In the current literature, solutions for multi-target tracking problem can be categorized into three main paradigms: the Joint Probabilistic Data Association (JPDA)~\cite{Fortmann1983Sonar}, Multiple Hypotheses Tracking (MHT)~\cite{Blackman2004Multiple}, and Random Finite Set (RFS)~\cite{Mahler2007Statistical}\footnote{We categorize these techniques considering the recent literature reviews presented in \cite{Mahler2007Statistical} and \cite{Vo2015Multitarget}. However, authors of \cite{Blackman1999Design} consider JPDA as a special case of MHT.}.  Among those, the RFS framework models the set of target states as a random variable instead of estimating measurement-to-track association hypothesis~\cite{Mahler2013Statistics102}. This formulation then allows the application of Bayesian recursion to produce the multi-target posterior density. 

In general, propagating a multi-target density is computationally expensive. Various filters only propagate the first moment of the multi-target density in the RFS paradigm, such as the Probability Hypothesis Density (PHD) filters~\cite{Mahler2003Multitarget,Vo2006Gaussian,Vo2007Analytic}. The recent introduction of labeled RFS~\cite{Vo2013Labeled} led to the development of filters capable of estimating the trajectories, such as the Generalized Labeled Multi-Bernoulli (GLMB) filters~\cite{Vo2014Labeled,Vo2019AMultiScan,Vo2019MultiSensor}, and  Labeled Multi-Bernoulli (LMB) approximation~\cite{Reuter2014TheLabeled}. The scalability of this approach has recently been demonstrated via its ability to track over one million targets~\cite{Beard2020ASolution}. Furthermore, labeled RFS filters can also be formulated to jointly track the targets and their ancestral information via a spawning model as in~\cite{Bryant2018AGeneralized,Nguyen2019Online, Nguyen2021Tracking}. Today, RFS-based filters have been applied to many fields ranging from space debris tracking~\cite{Wei2019MultiSensor, Jones2015Challenges}, crowd surveillance~\cite{Kim2019ALabeled, Ong2020ABayesian}, automation~\cite{Mullane2011ARandom, Nguyen2019Online} to cell tracking~\cite{Hadden2017Stem, Nguyen2021Tracking}.

Multi-sensor setting frequently appears in multi-target tracking applications.  Using multi-sensor framework allows the reduction of the system uncertainty, hence enhances the capability of the tracking algorithms to resolve the ambiguity of the target states. There are two architectures for multi-sensor multi-target tracking: distributed (decentralized) and centralized. 

In the distributed setting, information of targets is processed by individual sensor nodes then fused to form a posterior multi-target density. However, the correlation between the fused (from distributed sensor nodes) and the actual multi-sensor updated densities needs to be strong to warrant consistent tracking capability. In the RFS framework, the decentralized multi-sensor tracking problem has been addressed in~\cite{Uney2013Distributed, Battistelli2013Consensus} with PHD/CPHD filter, ~\cite{Guldogan2014Consensus, Yi2017Distributed,Yi2019Computationally} with Multi-Bernoulli filter, and in~\cite{Fantacci2018Robust, Li2017Robust,Li2018Computationally} with LMB/GLMB filter. 

In the centralized setting, measurements from all sensors are delivered to a central node for direct computation of the multi-target density. Solutions for this problem have been developed via PHD, CPHD filters in~\cite{Vo2004Tracking, Nannuru2016Multisensor}, Multi-Bernoulli filter in~\cite{Saucan2017AMultisensor}, or LMB filter in~\cite{Wang2018Centralized,Gostar2020Cooperative,Gostar2020Centralized,Panicker2020Tracking} . The iterated corrector method, which sequentially performs update overall sensors, is also widely used in practice~\cite{Vo2004Tracking, Pham2007Multiple}. Notably, the recent multi-sensor GLMB (MS-GLMB) filter~\cite{Vo2019MultiSensor} efficiently performs joint sensors update with linear complexity in the total number of measurements across the sensors and a quadratic complexity to the number of hypothesized targets. 

Information of targets detection probability and clutter rate (background information) influences the performance of multi-target filters while it is usually assumed to be known and constant. However, this assumption is strong and often does not hold in practice since these parameters are often time-varying in practical applications. Furthermore, the problem is worsened in multi-sensor tracking as background information must be provided for each sensor. For single-sensor systems, this problem has been addressed in~\cite{Beard2013Multitarget, Do2019Tracking, Mahler2011CPHD,Punchihewa2018Multiple}. However, it has not yet been tackled in the multi-sensor multi-target tracking framework.

This paper introduces a robust filter capable of estimating target trajectories with a centralized multi-sensor setting in an unknown background environment. Our main contribution is a systematic combination of multiple independent, robust CPHD filters~\cite{Mahler2011CPHD} to estimate clutter rate (each filter estimates the clutter rate of one sensor) and an efficient MS-GLMB~\cite{Vo2019MultiSensor} to handle the main filtering process. The detection probability of a target is augmented to its state for joint estimation \cite{Punchihewa2018Multiple}.

The structure of this paper is as follows. In Section~\ref{sec:Background}, we start with providing readers background information on multi-target tracking in the RFS paradigm. In Section~\ref{sec:The-proposed-filter}, we detail the formulation and implementation of our robust filter. Finally, in Section~\ref{sec:Numerical-study}, we demonstrate the performance of our filter in different tracking scenarios. Although this filter can handle different types of sensors, we focus on showing its performance with bearing-only sensors in this work.

\section{Background}\label{sec:Background}
This section briefly summarizes some fundamental concepts related to multi-target tracking in the RFS framework.
\subsection{The labeled RFS}
An RFS is a random variable defined on the space of sets with a finite number of elements. The elements of an RFS are random and unordered, which allows it to describe the multi-target state naturally. A labeled RFS is essentially the RFS, in which each element has a distinct label~\cite{Vo2013Labeled}. This framework allows the joint modelling of the target state and its identities; hence, filters based on labeled RFS can provide trajectories estimation.
For consistency, we adhere to the following notation scheme from~\cite{Vo2013Labeled} throughout our discussion. Specifically, the inner product of two functions $f$ and $g$ concerning their variable $x$ is defined as $\langle f,g\rangle \triangleq \int f\left(x\right)g\left(x\right)dx.$  For a given set $\mathcal{S}$, the class of all finite subsets of $\mathcal{S}$ is denoted by $\mathcal{F}(\mathcal{S})$. The Kronecker
delta function with arbitrary argument (integers, vectors, sets,
etc.) is given by,
\begin{equation}
   \delta_{\mathcal{S}}\left(X\right)=\begin{cases} 1, & X=\mathcal{S}\\
0, & X\neq\mathcal{S} \end{cases}.\end{equation}
The indicator function of the set $\mathcal{S}$ is defined as follows,
\begin{equation}
1_{\mathcal{S}}\left(X\right)=\begin{cases} 1, & X\subseteq\mathcal{S}\\
0, & \text{otherwise} \end{cases}.
\end{equation}

For a finite set $X$, the number of elements of $X$, or its cardinality, is denoted by $|X|$, and  the set exponential is defined as $\left[f\left(\cdot\right)\right]^{X}=\prod_{x\in X}f\left(x\right),$ with $f^\emptyset=1$.
 We denote the single-target state by lower-case letters and the multi-target states by upper-case letters, i.e. $x$ and $X$. We differentiate the unlabeled target state with the labeled version by standard-faced and bold-faced letters, i.e. $X$ and $\bm{X}$. Blackboard upper case letters (e.g. $\mathbb{X}, \mathbb{L}, \mathbb{Z}$) are used to denote spaces. 

\begin{figure}
        \centering
            \includegraphics[width=3in]{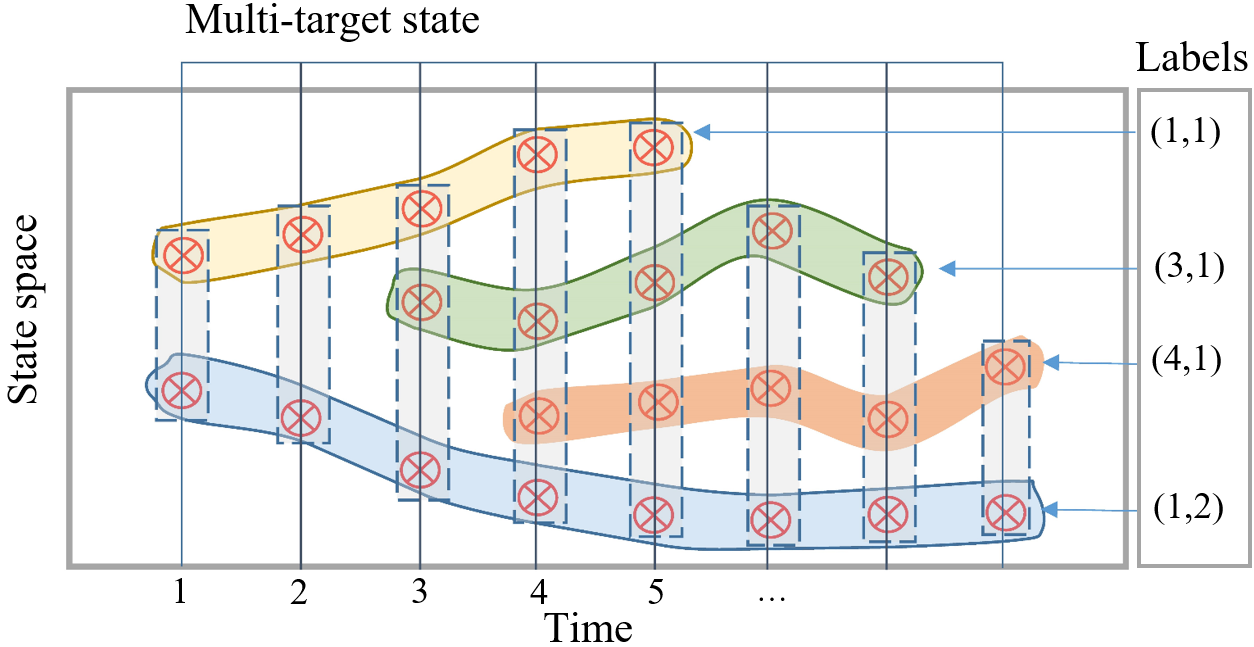}
        \caption{Evolution of the labeled multi-target state over time \cite{Vo2014Labeled}.}
        \label{figure 1}
    \end{figure}

Each labeled target state $\bm{x}$ contains an unlabeled kinematic state $x\in\mathbb{X}$ augmented with a unique label $\ell\in\mathbb{L}$, hence $\bm{x}=\left(x,\ell\right)\in\mathbb{X}\times\mathbb{L}$. Each distinct label $\ell\in\mathbb{L}$ at time $k$ consists of two components: time of birth, $t_{b}\leq k$, and a unique index $i$ to differentiate targets born at the same time. We use subscript `+' to denote the next time step. The birth labels at time $k+1$ belong to the label space $\mathbb{B}_{+}=\left\{ \left(k+1,i\right):i\in\mathbb{N}\right\}$, and hence $\mathbb{L\cap B_{+}=\emptyset}$. The labels space at time $k+1$ is then $\mathbb{L_{+}=L\uplus B_{+}}$ ($\uplus$ denotes the disjoint union operator). Fig. \ref{figure 1} illustrates the evolution of a labeled multi-target state over time  using the mentioned labeling convention.

The distinct label indicator~\cite{Vo2013Labeled} is given as,
\begin{equation} \label{distinct label indicator} \Delta\left(\bm{X}\right)=\delta_{|\bm{X}|}\left(\left|\mathcal{L}\left(\bm{X}\right)\right|\right), \end{equation}
where $\mathcal{L}:\mathbb{X\times L\rightarrow L}$ is a mapping from a labeled RFS to the labels, which satisfies the projection 
$\mathcal{L}(x,\ell)=\ell$. A labeled RFS $\bm{X}$ has distinct labels if $\Delta\left(\bm{X}\right)=1$.

The integral of a function $\bm{f} :\mathcal{F}\left(\mathbb{X\times L}\right)\rightarrow\mathbb{R}$ is given by~\cite{Vo2013Labeled}
    \begin{equation}
        \int \bm{f}\left(\bm{X}\right)\delta \bm{X}=\sum_{i=0}^{\infty}\frac{1}{i!}\sum_{\left(\ell_{1},\ldots,\ell_{i}\right)\in\mathbb{L}^{i}}\int_{\mathbb{X}^{i}}\bm{f}(\{ \left(x_{1},\ell_{1}\right), \ldots,\left(x_{i},\ell_{i}\right)\} )d\left(x_{1},\ldots,x_{i}\right).
   \label{eq:FISST integration} \end{equation}
\subsection{RFS multi-target filtering }
\subsubsection{The Bayesian recursion}

In Bayesian paradigm, given the multi-target transition density $\bm{f}_{+}\left(\bm{X}_{+}|\bm{X}\right)$
and the multi-target likelihood function $g\left(Z_{+}|\bm{X}_{+}\right)$,
the (labeled) multi-target probability density function\footnote{This is not a probability density function but is equivalent to one
as shown in~\cite{Vo2005Sequential}. Hence, with a slight abuse
of terminology, we regard this function as a probability density function.} is propagated via the Bayesian recursion as~\cite{Mahler2007Statistical},
    \begin{align} \bm{\pi}\left(\bm{X}_{+}\right) =\int\bm{f}_{+}\left(\bm{X}_{+}|\bm{X}\right)\bm{\pi}\left(\bm{X}\right)\delta\bm{X}, \label{eq:prior recursive} \\
    \bm{\pi}_{+}\left(\bm{X}_{+}|Z_{+}\right) =\frac{g_{+}\left(Z_{+}|\bm{X}_{+}\right)\bm{\pi}\left(\bm{X}_{+}\right)}{\int g_{+}\left(Z_{+}|\bm{X}\right)\bm{\pi}\left(\bm{X}\right)\delta\bm{X}}, \label{eq:posterior recursive} \end{align}
where the integrals are the set integral defined in Eq.~\eqref{eq:FISST integration}.
In the above expression, the dependence on previous measurements of
the prior density is omitted for compactness.

\subsection{The multi-target transition model}

For a current multi-target state $\bm{X}\subseteq\mathbb{X}\times\mathbb{L}$, at the
next time step, a target $(x,\ell)\in\bm{X}$ either continues to
exist in the sensor field of view with the probability $p_{S}(x,\ell)$, or it disappears
with the probability $q_{S}(x,\ell)=1-p_{S}(x,\ell)$. If it exists,
its state is predicted via a single target transition density of the form $f_{S}(x_{+}|x,\ell)\delta_{\ell}(\ell_{+})$
where $f_{S}(x_{+}|x,\ell)$ is the spatial transition density and
$\delta_{\ell}(\ell_{+})$ implies the old label is retained. The multi-target transition density from set $\bm{X}$ to the next
time step surviving targets set $\bm{X}_{S+}$ can be written as~\cite{Vo2013Labeled},
\begin{equation}
\bm{f}_{S+}\left(\bm{X}_{S+}|\bm{X}\right)=\Delta\left(\bm{X}_{S+}\right)\Delta\left(\bm{X}\right)1_{\mathcal{L}\left(\bm{X}\right)}\left(\mathcal{L}\left(\bm{X}_{S+}\right)\right)\left[\Phi_{S+}\left(\bm{X}_{S+}|\cdot\right)\right]^{\bm{X}},\label{eq:survival distribution}
\end{equation}
where
\begin{equation}
\Phi_{S+}\left(\bm{X}_{S+}|x,\ell\right)=\sum_{\left(x_{+},\ell_{+}\right)\in\bm{X}_{S+}}\delta_{\ell}\left(\ell_{+}\right)p_{S}\left(x,\ell\right)f_{S+}\left(x_{+}|x,\ell\right)+\left[1-1_{\mathcal{L}\left(\bm{X}_{S+}\right)}\left(\ell\right)\right]q_{S}\left(x,\ell\right).\label{eq:Phi_S}
\end{equation}
  
In addition, new targets can also spontaneously appear in the tracking
region at each time step. Let $\bm{X}_{B+}$ denote the multi-target
state of those newborn targets with $\mathcal{L}(\bm{X}_{B+})\in\mathbb{B}_{+}$,
the set of new births can be modeled with an LMB RFS as~\cite{Vo2013Labeled,Vo2014Labeled},
\begin{equation}
\bm{f}_{B+}\left(\bm{X}_{B+}\right)=\Delta\left(\bm{X}_{B+}\right)\omega_{B}\left(\mathcal{L}\left(\bm{X}_{B+}\right)\right)\left[p_{B+}\right]^{\bm{X}_{B+}},\label{eq:birth distribution}
\end{equation}
where
\begin{equation}
\omega_{B}\left(L\right)=\left[1-r_{B+}\right]^{\mathbb{B}_{+}-L}1_{\mathbb{B}+}\left(L\right)\left[r_{B+}\right]^{L},
\end{equation}
$r_{B+}(\ell)$ is the birth probability of target labeled $\ell$
and $p_{B+}(x,\ell)$ is its spatial distribution.

Due to the independence between surviving and newborn targets, the
multi-target transition kernel can be written as, 
\begin{equation}
\bm{f}_{+}\left(\bm{X}_{+}|\bm{X}\right)=\bm{f}_{S_{+}}\left(\bm{X}_{S+}|\bm{X}\right)\bm{f}_{B+}\left(\bm{X}_{B+}\right),\label{transition kernel without spawning}
\end{equation}
where $\bm{X}_{S+}=\bm{X}_{+}\cap\left(\mathbb{X\times L}\right)$,
and $\bm{X}_{B+}=\bm{X}_{+}\cap\left(\mathbb{X\times B_{+}}\right)$.

\subsubsection{The single-sensor multi-target likelihood model}

At time $k$, each target $(x,\ell)\in\bm{X}$ can generate a measurement
$z\in Z$ with the detection probability of $p_{D}(x,\ell)$ or it
is misdetected with the probability $q_{D}(x,\ell)=1-p_{D}(x,\ell)$.
The likelihood of target $(x,\ell)$ generates a measurement $z$ is
$g(z|x,\ell)$. In addition, due to sensor imperfections and environmental
conditions, clutter can also be included in the observation set.
The clutter set is modelled with Poisson RFS, with the state of clutter targets is assumed
to be uniformly distributed over the state space. Denoting $\mathbb{Z}$
as the measurement space, given the set $Z=z_{1:|Z|}\in\mathbb{Z}$
and the multi-target state $\bm{X}$ at the current time step $k$, the
multi-target likelihood function is given as~\cite{Vo2013Labeled}
\begin{equation}
g(Z|\bm{X})\propto\sum_{\theta\in\Theta(\mathcal{L}(\bm{X}))}\prod_{(x,\ell)\bm{X}}\Psi_{Z}^{(\theta(\ell))}(x,\ell),\label{multitarget likelihood function}
\end{equation}
where $\Theta$ is the set of all positive $1$-$1$ labels to measurement
indices association maps ($\theta:\mathbb{L\rightarrow}\left\{ 0:|Z|\right\} $)
with the zero measurement index indicates miss-detection. The positive 1-1 condition
implies each measurement can only be assigned to at most one label.
The function $\Psi$ is given by 
\begin{equation}
\Psi_{Z}^{(\theta(\ell))}(x,\ell)=\begin{cases}
\frac{P_{D}(x,\ell)g(z_{\theta(\ell)}|x,\ell)}{\kappa(z_{(\theta(\ell)})}, & \theta(\ell)>0\\
1-P_{D}(x,\ell), & \theta(\ell)=0,
\end{cases},\label{psiZ}
\end{equation}
where $\kappa$ is the Poisson clutter intensity.

\subsection{The GLMB filter}\label{subsec:delta_GLMB} 

GLMB filter is an exact solution for the Bayes optimal multi-target tracking problem~\cite{Mahler2019Exact}. Given its formulation on labeled RFS, GLMB filter estimates target states and their labels which, in turn, allows the estimation of trajectories.
The filter first assumes the probability density function of the multi-target state at time $k$ be given in the GLMB form, i.e.
\begin{equation}
    \bm{\pi}\left(\bm{X}\right)=\Delta\left(\bm{X}\right)\sum_{\left(I,\xi\right)\in\mathcal{F}\left(\mathbb{L}\right)\times\Xi}\omega^{\left(I,\xi\right)}\delta_{I}\left(\mathcal{L}\left(\bm{X}\right)\right)\left[p^{\left(\xi\right)}\right]^{\bm{X}},    \label{eq: GLMB prior}
\end{equation}
where $I$ represents a set of labels, and $\xi\in\Xi$ represents a history of association maps up to time $k$. Each $p^{\left(\xi\right)}(\cdot,\ell)$ represents a spatial distribution of a single target on $\mathbb{X}$ (with $\int p^{\left(\xi\right)}\left(x,\ell\right)dx=1$), and each non-negative weight $\omega^{\left(I,\xi\right)}$ satisfies,
\begin{equation}
    \sum_{I\in\mathcal{F}\left(\mathbb{L}\right)}\sum_{\xi\in\Xi}\omega^{\left(I,\xi\right)}\left(L\right)=1. \label{standard prior weight}
\end{equation}

Via Bayes recursion with the multi-target transition model in \eqref{transition kernel without spawning} and likelihood function in \eqref{multitarget likelihood function}, the prior GLMB density is propagated through time \cite{Vo2014Labeled}. A more efficient implementation that combines the prediction and update steps into one single filtering step is also presented in \cite{Vo2017AnEfficient}.

\section{The robust MS-GLMB filter} \label{sec:The-proposed-filter} 
\begin{figure}[H] 
\centering \includegraphics[width=3.5in]{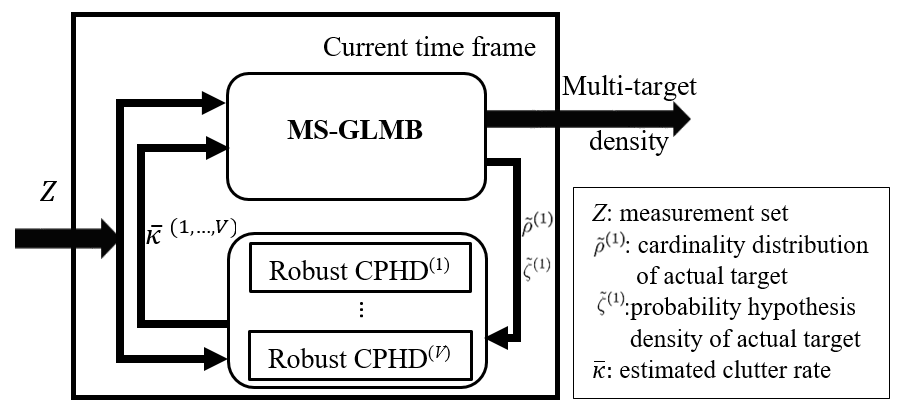} 
\caption{The proposed structure  for the robust MS-GLMB filter.}
\label{figure_2} 
\end{figure}

Our proposed robust MS-GLMB filter  exploits the strengths of established filters in the literature to track multiple targets without prior knowledge of the detection profile and clutter rates. In particular, for each sensor, we implement an independent, robust CPHD filter \cite{Mahler2011CPHD} to estimate its clutter rate from the PHD and cardinality distribution of targets set obtained from the GLMB density at the previous time step. The estimated clutter rate is then bootstrapped into the MS-GLMB filter \cite{Vo2019MultiSensor} for the main filtering process. The detection probability of a target is augmented to the filtering state space of the filters for estimation \cite{Punchihewa2018Multiple}. The schematic of the algorithm is given in Fig.~\ref{figure_2}.

\subsection{ The multi-target system modelling}
\subsubsection{The robust CPHD filter}\label{subsec:model-CPHD}
The CPHD filter is a low-cost multi-target  filter built on the premise of unlabeled RFS. In this filter, the multi-target density is approximated by its first moment (PHD) and the cardinality distribution. While being more accurate than the PHD filter, the CPHD filter  has a complexity of $\mathcal{O}\left(\left|Z\right|^{3}\right)$ or even $\mathcal{O}\left(\left|Z\right|^{2} \log^2(|Z|)\right) $  \cite{Vo2007Analytic}. In~\cite{Mahler2011CPHD}, the CPHD filter is  adapted to jointly estimate the clutter rate and detection probability (the robust CPHD) online.  Note that, the clutter rate of each sensor can also be estimated using data association method (with a single-sensor GLMB filter) as proposed in \cite{Punchihewa2018Multiple}. However, for each prior component of the single-sensor GLMB density, it incurs  a complexity of $\mathcal{O}\left(T\left|Z\right|^{3}\right)$ (where $T$ is the number of request components) to sample for significant components. Such complexity prevents the application of the data association approach in scenarios with a high number of sensors. Hence, the robust CPHD filter provides a good trade-off between accuracy and computational load in estimating the sensor clutter rate.

Following~\cite{Mahler2011CPHD}, the clutter and actual targets are modelled as two different types of targets. The detection probability is augmented to the target state for joint estimation. Denoting the state spaces for actual and clutter targets as $\mathbb{X}^{(1)}$ and $\mathbb{X}^{(0)}$, the hybrid state space is then given by~\cite{Mahler2011CPHD}:
\begin{equation}
    \mathbb{X}^{(h)}=\left(\mathbb{X}^{(1)}\times\mathbb{X}^{(\Delta)}\right)\uplus\left(\mathbb{X}^{(0)}\times\mathbb{X}^{(\Delta)}\right),
    \label{hybrid_ss-1}
\end{equation}
where $\mathbb{X}^{(\Delta)}=[0,1]$ is the space of unknown detection probability.

For consistency, we use the superscripts $^{(1)},^{(0)}$ and $^{(h)}$ to denote functions or variables on space of actual, clutter, and hybrid targets, respectively. 

For the $v^{th}$ sensor, at the next time step, a target survives with the probability $p_{S}^{(v,h)}(x^{(h)})$ and has the transition density of $f_{+}^{(v)}(x_{+}^{(h)}|x^{(h)})$ or being terminated with the probability $1-p_{S}^{(v,h)}(x^{(h)})$. The survival probability and transition density are given respectively as~\cite{Mahler2011CPHD}:
\begin{equation}
    p_{S}^{(v,h)}(x^{(h)})=\begin{cases}
p_{S}^{(1)}(x), & x^{(h)}\in\mathbb{X}^{(1)}\times\mathbb{X}^{(\Delta)}\\
p_{S}^{(v,0)}, & x^{(h)}\in\mathbb{X}^{(0)}\times\mathbb{X}^{(\Delta)}
\end{cases},
\end{equation}
\begin{equation}
    f_{+}^{(v)}(x_{+}^{(h)}|x^{(h)})=\begin{cases}
f_{+}^{(1)}(x_{+}|x)f_{\Delta+}^{(v)}(a_{+}|a), & x_{+}^{(h)}=(x_{+},a_{+}),x^{(h)}=(x,a)\in\mathbb{X}^{(1)}\times\mathbb{X}^{(\Delta)}\\
f_{+}^{(v,0)}(c_{+}|c), & x_{+}^{(h)}=(c_{+},b_{+}),x^{(h)}=(c,b)\in\mathbb{X}^{(0)}\times\mathbb{X}^{(\Delta)}\\
0, & \textrm{otherwise}
\end{cases}.
\end{equation}

Given a target with the state $x^{(h)}$ defined on the hybrid state-space, it can generate a measurement $z\in Z$ with the probability $p_{D}^{(h)}(x^{(h)})$ and the likelihood $g^{(v,h)}(z|x^{(h)})$, or being miss-detected with the probability of $1-p_{D}^{(h)}(x^{(h)})$. The detection probability and the likelihood of observing $z$ are given respectively as~\cite{Mahler2011CPHD}: 
\begin{align}
    p_{D}^{(h)}(x^{(h)})&	=	\begin{cases}
a, & \qquad\quad x^{(h)}=(x,a)\in\mathbb{X}^{(1)}\times\mathbb{X}^{(\Delta)}\\
b, & \qquad \quad x^{(h)}=(c,b)\in\mathbb{X}^{(0)}\times\mathbb{X}^{(\Delta)}
\end{cases},\\
g^{(v,h)}(z|x^{(h)})&	=	\begin{cases}
g^{(v)}(z|x) & x^{(h)}=(x,a)\in\mathbb{X}^{(1)}\times\mathbb{X}^{(\Delta)}\\
\mu^{(v)}(z) & x^{(h)}=(c,b)\in\mathbb{X}^{(0)}\times\mathbb{X}^{(\Delta)}
\end{cases}.
\end{align}
\subsubsection{The MS-GLMB filter}\label{subsec:model-MS-GLMB}
The MS-GLMB filter~\cite{Vo2019MultiSensor} is an extension of the efficient GLMB filter~\cite{Vo2017AnEfficient} to the multi-sensor framework. Given its labeled RFS formulation, the MS-GLMB filter propagates the labeled multi-target density hence providing trajectory estimates. In this work, the MS-GLMB filter is formulated to carry out the main filtering process. The sensor clutter rates are bootstrapped from the robust CPHD filters.

Assuming the detection probability of a target on different sensors is independent, we can write the state distribution
of a target with label $\text{\ensuremath{\ell}}$ and association
history $\xi$ as $p^{(\xi)}(x,\alpha,\ell)=p^{(\xi)}(x,\ell)\prod_{v=1}^{V}p^{(\xi)}(\alpha_{v})$.
Given the inclusion of the detection probability, the state transition
model in Eq.~\eqref{eq:Phi_S} can be rewritten as:
\begin{equation}
\begin{aligned}
    \Phi_{S+}\left(\bm{X}_{S+}|x,\alpha,\ell\right)	=&	\sum_{\left(x_{+},\ell_{+}\right)\in\bm{X}_{S+}}\delta_{\ell}\left(\ell_{+}\right)p_{S}\left(x,\ell\right)f_{S+}\left(x_{+}|x,\ell\right)\\
		& \times\prod_{v=1}^{V}f_{\Delta+}^{(v)}\left(\alpha_{v+}|\alpha_{v}\right)+\left[1-1_{\mathcal{L}\left(\bm{X}_{S+}\right)}\left(\ell\right)\right]q_{S}\left(x,\ell\right),
\end{aligned} \label{eq:transition_with_pD}
\end{equation}
where $f_{v+}^{(\Delta)}\left(\cdot|\cdot\right)$is the state transition density of the detection probability of the target on the $v^{th}$ sensor.

For multi-sensor multi-target tracking, we follow~\cite{Vo2019MultiSensor} to extend the likelihood in Eq.~\eqref{multitarget likelihood function} to multiple sensors case. Specifically, given the multi-target state $\bm{X}$ and the set of measurements $Z^{(v)}={z_{1:|Z^{(v)}|}^{(v)}}\in\mathbb{Z}$ produced by sensor $v^{th}$, each observation of a target $(x,\alpha,\ell)\in\bm{X}$ on this sensor is either a measurement $z_{j}^{(v)}\in Z^{(v)}$ with the detection probability of $\alpha_{v}$ and the likelihood of $g^{(v)}(z^{(v)}|x,\ell)$ or empty (miss-detected) with the probability $1-\alpha_{v}$. Furthermore, the set $Z^{(v)}$ may also contain measurements that are not generated by any targets and the set of those false measurements is modeled by Poisson RFS with the rate (intensity) of $\kappa^{(v)}$ (estimated by the robust CPHD filter). The multi-target likelihood of this sensor can be written as ~\cite{Vo2013Labeled,Vo2014Labeled},
\begin{equation}
    g^{(v)}\left(Z^{(v)}|\bm{X}\right)\propto\sum_{\theta^{(v)}\in\Theta^{(v)}}1_{\Theta^{(v)}(\mathcal{L}(\bm{X}))}\left(\theta^{(v)}\right)\left[\psi_{Z^{(v)}}^{\left(v,\theta^{(v)}\circ\mathcal{L}(\cdot)\right)}\left(\cdot\right)\right]^{\bm{X}},
    \label{standard multitarget observation model}
\end{equation}
    where 
    \begin{equation}
        \psi_{\{z_{1:M^{(v)}}\}}^{\left(v,j\right)}\left(x,\alpha,\ell\right)=\begin{cases}
\frac{\alpha_{v}g^{(v)}(z_j|(x,\ell))}{\kappa^{(v)}(z_{j})} & j=1:M^{(v)}\\
1-\alpha_{v} & j=0.
\end{cases}
\label{eq:Psi_Z}
    \end{equation}
    Notations are defined as the same as those in Eq.~\eqref{multitarget likelihood function}  with the additional superscript $^{(v)}$ to indicate the sensor index. 
    
Via the assumption that all sensors are conditionally independent and the following abbreviations: 
      \begin{align}
        Z&\triangleq\left(Z^{(1)},\ldots,Z^{(V)}\right), \\ \Theta&\triangleq\Theta^{(1)}\times\cdots\times\Theta^{(V)},\\
\Theta(I)&\triangleq\Theta^{(1)}(I)\times\cdots\times\Theta^{(V)}(I),\\ \theta&\triangleq\left(\theta^{(1)},\ldots,\theta^{(V)}\right),\\
1_{\Theta(I)}(\theta)&\triangleq\prod_{v=1}^{V}1_{\Theta^{(v)}(I)}\left(\theta^{(v)}\right), \\
\psi_{Z}^{\left(j^{(1)},\ldots,j^{(V)}\right)}&\triangleq\prod_{v=1}^{V}\psi_{Z^{v}}^{(v,j^{(v)})}(x,\alpha,\ell),
 \end{align}
 the multi-sensor multi-target likelihood can be written simply as:
\begin{equation}
    g(Z|\bm{X})	=\prod_{v=1}^{V}g^{(v)}\left(Z^{(v)}|\bm{X}\right)\propto\sum_{\theta\in\Theta}1_{\Theta(\mathcal{L}(\bm{X}))}(\theta)\left[\psi_{Z}^{(\theta\circ\mathcal{L}(\cdot))}(\cdot)\right]^{\bm{X}}.
    \label{Multi-sensorlikelihoodfunct}
\end{equation}
Note that, as all the terms $\theta^{(i)}$ ($i=1, \ldots,V$) are positive 1-1, the multi-sensor $\textit{extended association map }  \theta$ is also positive 1-1.

Remark. For tracking scenarios  involving target spawning (i.e., a new target is born from an existing target), the target spawning model presented in \cite{Bryant2018AGeneralized} can be used in place of the dynamic model in Eq. (\ref{eq:survival distribution}) for the MS-GLMB filter. For the CPHD filter, the standard dynamic model (without target spawning) is sufficient given its unlabeled formulation \cite{Mahler2014Advances} (Chapter 1, Section 1.1.2). For tractability, a proposal density can be constructed by replacing the single-sensor likelihood in Eq. (43) of \cite{Bryant2018AGeneralized} with the multi-sensor likelihood in Eq. (\ref{Multi-sensorlikelihoodfunct}). Hence, our method is readily extended to this tracking scenario. However, the inclusion of the spawning model naturally increases the complexity of the filter due to additional operations (approximation, marginalization) need to be performed.

\subsection{Clutter rate estimation with the robust CPHD filter}\label{subsec:prob_dect_clutt}

For the $v^{th}$ sensor, the recursion of the robust CPHD filter at each time step starts with the prior PHD and cardinality distributions of targets on the hybrid state space. Note that we also have the prior GLMB density of the labeled targets state (from the MS-GLMB filter). Since the MS-GLMB filter combines measurements from all sensors, its contained information is more accurate than that of the current $v^{th}$ robust CPHD filter (which is only updated by measurements from the $v^{th}$ sensor). Basing on that fact, instead of using the prior PHD and cardinality from the current robust CPHD filter, we use prior information from the GLMB density to predict the PHD and cardinality distribution at the next time step. Specifically, given the GLMB prior as in Eq.~\eqref{eq: GLMB prior}, the PHD and cardinality distribution of an actual target can be computed, respectively as 
\begin{align}
    \tilde{\zeta}^{\left(1\right)}\left(x,\alpha_{v}\right)	=&	\sum_{I,\xi}\sum_{\ell\in I}\omega^{(I,\xi)}p^{(\xi)\!}\left(x,\ell\right)p^{(\xi)\!}\left(\alpha_{v}\right)\prod_{i\in\{1:V\}-\{v\}}\int p^{(\xi)\!}\left(\alpha_{i}\right)d\alpha_{i},\label{eq:GLMB-PHD}\\
\tilde{\rho}^{(1)}(n)	=&	\sum_{I,\xi}\delta_{n}\left(\left|I\right|\right)\omega^{(I,\xi)}\label{eq:GLMB-card}.
\end{align}
The cardinality distribution of targets on the hybrid state space is then $\tilde{\rho}^{(h)}=\tilde{\rho}^{(1)}\ast\rho^{(0)}$ (`$\ast$' denotes convolution operation).

Given Eq.~\eqref{eq:GLMB-PHD}, Eq.~\eqref{eq:GLMB-card} and the priors information of the current sensor clutter targets, the robust CPHD prediction can be carried out as~\cite{Mahler2011CPHD}:
\begin{align}
    \zeta_{+}^{\left(v,1\right)}\left(x_{+},a_{+}\right)	=&	\gamma^{(1)}(x_{+},a_{+})+\int\int p_{S}^{(1)}(x)f_{+}^{(1)}(x_{+}|x)f_{\Delta+}^{(v)}(a_{+}|\alpha_{v})\tilde{\zeta}^{\left(1\right)}\left(x,\alpha_{v}\right)d\alpha_{v}\thinspace dx,\\
\zeta_{+}^{\left(v,0\right)}\left(b\right)	=&	\gamma^{(v,0)}(b)+p_{S}^{(v,0)}\zeta^{\left(v,0\right)}\left(b\right),\\
\rho_{+}^{(v,h)}(n)	=&	\sum_{j=0}^{n}\rho_{+}^{(v,h)}(n-j)\sum_{i=j}^{\infty}C_{j}^{i}\tilde{\rho}^{(h)}(i)(1-\phi_{v})^{i-j}\phi_{v}^{j},\\
\phi_{v}	=&	\frac{\left\langle \tilde{\zeta}^{\left(1\right)},p_{S}^{(1)}\right\rangle +\left\langle \zeta^{\left(v,0\right)},p_{S}^{(v,0)}\right\rangle }{\left\langle 1,\tilde{\zeta}^{\left(1\right)}\right\rangle +\left\langle 1,\zeta^{\left(v,0\right)}\right\rangle },
\end{align}
where $C_{j}^{i}$ is the binomial coefficient, i.e., $C_{j}^{i}=\frac{i!}{j!(i-j)!} $.

Given the measurements set $Z_{+}^{(v)}$, the updated PHD and cardinality distribution can be written as~\cite{Mahler2011CPHD}:
\begin{equation}
\begin{aligned}
\zeta_{+}^{\left(v,1\right)}&\left(x_{+},a_{+}|Z_{+}^{(v)}\right)=\zeta_{+}^{\left(v,1\right)}\left(x_{+},a_{+}\right)\times\\
&\left[\frac{\left(1-a_{+}\right)\times\frac{\langle\Gamma_{Z_{+}^{(v)}}^{\left(v,1\right)}\left[\zeta_{+}^{\left(v,h\right)},Z_{+}^{(v)}\right],\rho_{+}^{\left(v,h\right)}\rangle}{\langle\Gamma_{Z_{+}^{(v)}}^{\left(v,0\right)}\left[\zeta_{+}^{\left(v,h\right)},Z_{+}^{(v)}\right],\rho_{+}^{\left(v,h\right)}\rangle}}{\langle1,\zeta_{+}^{\left(v,1\right)}\rangle+\langle1,\zeta_{+}^{\left(v,0\right)}\rangle}+\sum_{z\in Z_{+}^{(v)}}\frac{a_{+}\times g^{(v,1)}\left(z|x_{+}\right)}{\langle\zeta_{+}^{\left(v,0\right)},p_{D_{+}}^{\left(v,0\right)}\mu^{(v)}\rangle+\langle\zeta_{+}^{\left(v,1\right)},p_{D_{+}}^{\left(v,1\right)}g^{\left(v,1\right)}\left(z|\cdot\right)\rangle}\right],
 \end{aligned}   
\end{equation}
\begin{equation}
    \begin{aligned}
    \zeta_{+}^{\left(v,0\right)}&\left(b_{+}|Z_{+}^{(v)}\right)=\zeta_{+}^{(v,0)}\left(b_{+}\right)\times\\
&\left[\frac{\left(1-b_{+}\right)\times\frac{\langle\Gamma_{Z_{+}^{(v)}}^{\left(v,1\right)}\left[\zeta_{+}^{\left(v,h\right)},Z_{+}^{(v)}\right],\rho_{+}^{\left(v,h\right)}\rangle}{\langle\Gamma_{+}^{\left(v,0\right)}\left[\zeta_{+}^{\left(v,h\right)},Z_{+}^{(v)}\right],\rho_{+}^{\left(v,h\right)}\rangle}}{\langle1,\zeta_{+}^{\left(v,1\right)}\rangle+\langle1,\zeta_{+}^{\left(v,0\right)}\rangle}+\sum_{z\in Z_{+}^{(s)}}\frac{b_{+}\times\mu^{(v)}(z)}{\langle\zeta_{+}^{\left(v,0\right)},p_{D_{+}}^{\left(v,0\right)}\mu^{(v)}\rangle+\langle\zeta_{+}^{\left(v,1\right)},p_{D+}^{\left(v,1\right)}g^{\left(v,1\right)}\left(z|\cdot\right)\rangle}\right],\\
    \end{aligned}
\end{equation}
\begin{equation}
    \rho_{+}^{(v,h)}(n)=\begin{cases}
0, & n<|Z_{+}^{(v)}|\\
\frac{\rho(n)\cdot\Gamma^{(v,0)}[\zeta_{+}Z_{+}^{(v)}](n)}{\left\langle \rho_{+}^{(v,h)},\Gamma_{+}^{\left(v,0\right)}\right\rangle }, & n\geq|Z_{+}^{(v)}|
\end{cases},
\end{equation}
where 
\begin{equation}
    \Gamma_{+}^{\left(v,u\right)}\left[\zeta_{+}^{\left(v,h\right)}Z_{+}^{(v)}\right]\left(n\right)	=\begin{cases}
0, & n<|Z_{+}^{(v)}|+u\\
P_{|Z_{+}^{(v)}|+u}^{\left(n\right)}\Phi_{v+}^{\left(n-\left(|Z_{+}^{(v)}|+u\right)\right)}, & n\geq|Z_{+}^{(v)}|+u
\end{cases},
\end{equation}
\begin{align}
    \Phi_{v+}=& 1-\frac{\langle\zeta_{+}^{\left(v,1\right)},p_{D_{+}}^{\left(v,1\right)}\rangle+\langle\zeta_{+}^{\left(v,0\right)},p_{D+}^{\left(v,0\right)}\rangle}{\langle1,\zeta_{+}^{\left(v,1\right)}\rangle+\langle1,\zeta_{+}^{\left(v,0\right)}\rangle},\\ p_{D+}^{\left(v,1\right)}(x,a)=&a,\\ p_{D+}^{\left(v,0\right)}(b)=&b,
\end{align}
and $P_{j}^{n}$ is the permutation coefficient, i.e., $P_{j}^{n}=\frac{n!}{(n-j)!}$ .

In this work, we model the kinematic state and detection probability of targets with Gaussian and beta distributions, respectively. The analytic implementation can be found in Proposition 13 and 14 of~\cite{Mahler2011CPHD}. For clutter rate estimation, given the PHD of clutter targets is written as a mixture of beta distributions, i.e.
\begin{equation}
    \zeta^{(v,0)}(b_{+})=\sum_{i=1}^{J^{(v,0)}}w_{i}^{(v,0)}\beta_{i}^{(v,0)}(b_{+}|s_{i}^{(v,0)},t_{i}^{(v,0)}),
\end{equation}
where $\beta(\cdot|s,t)$ denotes a beta distribution with mean $\frac{s}{(s+t)}$ and variance $\frac{st}{(s+t)^{2}(s+t+1)}$. The mean clutter rate for the $v^{th}$ sensor is computed as \cite{Mahler2011CPHD}
\begin{equation}
   \bar{\kappa}^{(v)}=\sum_{i=1}^{J^{(v,0)}}w_{i}^{(v,0)}\frac{s_{i}^{(v,0)}}{s_{i}^{(v,0)}+t_{i}^{(v,0)}}.\label{eq:est-clutter-rate}
\end{equation}
Practically, the presented clutter rate estimation step can be parallelized for $V$ sensors to reduce the computation time.

\subsection{Multi-target state filtering with the MS-GLMB filter}
Given the set of estimated clutter rates $\bar{\kappa}^{(1:V)}$ (see Eq.~\eqref{eq:est-clutter-rate}), the main multi-target filtering is carried out via the MS-GLMB filter.  Noting that the detection probability $\alpha$ of a target is augmented to its state for joint estimation as in \cite{Punchihewa2018Multiple}. For the initial GLMB density of the form in Eq.~\eqref{eq: GLMB prior}, the transition model in Eq.~\eqref{transition kernel without spawning}, Eq.~\eqref{eq:transition_with_pD} and the multi-sensor multi-target likelihood in Eq.~\eqref{multitarget likelihood function}, the filtering density at the next time step is given as:
\begin{equation}
    \bm{\pi}_{+}\left(\bm{X}_{+}|Z_{+}\right)\propto\Delta\left(\bm{X}_{+}\right)\sum_{I,\xi,I_{+},\theta_{+}}\omega^{\left(I,\xi\right)}\omega_{Z_{+}}^{\left(I,\xi,I_{+},\theta_{+}\right)}\delta_{I_{+}}\left[\mathcal{L}\left(\bm{X}_{+}\right)\right]\left[p_{Z_{+}}^{\left(\xi,\theta_{+}\right)}\right]^{\boldsymbol{X}_{+}},\label{GLMB posterior-1}
\end{equation}
where $I\in\mathcal{F}\left(\mathbb{L}\right),\xi\in\Xi,I_{+}\in\mathcal{F}\left(\mathbb{L}_{+}\right),\theta_{+}\in\Theta_{+}\left(I_{+}\right),$  and
\begin{align}
    \omega_{Z_{+}}^{\left(I,\xi,I_{+},\theta_{+}\right)}	=&	1_{\Theta_{+}\left(I_{+}\right)}\left(\theta_{+}\right)\left[1-\bar{P}_{S}^{\left(\xi\right)}\right]^{I-I_{+}}\left[\bar{P}_{S}^{\left(\xi\right)}\right]^{I\cap I_{+}}\left[1-r_{B,+}\right]^{\mathbb{B}_{+}-I_{+}}r_{B,+}^{\mathbb{B}_{+}\cap I_{+}}\left[\bar{\psi}_{Z_{+}}^{\left(\xi,\theta_{+}\right)}\right]^{I_{+}},\\
\bar{P}_{S}^{\left(\xi\right)}\left(\ell\right)	=&	\bigl\langle p^{\left(\xi\right)}\left(\cdot,\ell\right),P_{S}\left(\cdot,\ell\right)\bigr\rangle,\\
\bar{\psi}_{Z_{+}}^{\left(\xi,\theta_{+}\right)}\left(\ell_{+}\right)	=&	\int\bar{p}_{+}^{\left(\xi\right)}\left(x_{+},\ell_{+}\right)\prod_{v=1}^{V}p^{(\xi)}(\alpha_{v})\psi_{Z_{+}}^{\left(\theta_{+}\left(\ell_{+}\right)\right)}\left(x_{+},\alpha_{+},\ell_{+}\right)dx_{+}d\alpha_{1:V},\\
p_{S}\left(x_{+},\alpha_{+},\ell_{+}\right)	=&	\int P_{S}\left(x,\ell_{+\!}\right)f_{S+}\left(x_{+}|x,\ell_{+}\right)p^{\left(\xi\right)}\left(x,\ell_{+}\right)dx\times\prod_{v=1}^{V}\int p^{(\xi)}(\alpha_{v})f_{\Delta+}^{(v)}(\alpha_{v+}|\alpha_{v})d\alpha_{v},\\
\bar{p}_{+}^{\left(\xi\right)}\left(x_{+},\alpha_{+},\ell_{+}\right)	=&	1_{\mathbb{L}}\left(\ell_{+}\right)\frac{p_{S}\left(x_{+},\alpha_{+},\ell_{+}\right)}{\bar{P}_{S}^{\left(\xi\right)}\left(\ell_{+}\right)}+1_{\mathbb{B}_{+}}\left(\ell_{+}\right)p_{B,+}\left(x_{+},\ell_{+}\right),\\
p_{Z_{+}}^{\left(\xi,\theta_{+}\right)}\left(x_{+},\alpha_{+},\ell_{+}\right)	=&	\frac{\bar{p}_{+}^{\left(\xi\right)}\left(x_{+},\alpha_{+},\ell_{+}\right)\psi_{Z_{+}}^{\left(\theta_{+}\left(\ell_{+}\right)\right)}\left(x_{+},\alpha_{+}\ell_{+}\right)}{\bar{\psi}_{Z_{+}}^{\left(\xi,\theta_{+}\right)}\left(\ell_{+}\right)}.
\end{align}

The number of components in the GLMB filtering density grows exponentially over time. Hence, it needs to be truncated to keep the filter tractable. To select significant components, one approach is to formulate the multi-dimensional assignment on the extended association maps. However, solving this problem is NP-hard and intractable given a high number of sensors. On the other hand, significant components can be efficiently sampled from a stationary distribution via the Gibbs sampler. However, this approach is also intractable with a high number of sensors due to the amount of memory required to store the high-dimensional distribution. In this work, minimally-Markovian (between sensors) on the stationary distribution is assumed~\cite{Vo2019MultiSensor}, which allows a significant reduction of computation time and memory usage. Hence it allows the filter to operate with a high number of sensors in an online fashion. The filtering procedure is given in Alg. 2 of~\cite{Vo2019MultiSensor}.

In this filter, we model the kinematic with the Gaussian distribution, and the detection probability of each sensor with independent beta distribution. The procedures to predict and update each beta distribution are described in~\cite{Mahler2011CPHD}.

The pseudo-code in Alg.~\ref{alg:robust_MSGLMB_pseudo} lays out the implementation of our algorithm. The modelCPHD and modelGLMB classes contain the corresponding transition and likelihood models for the CPHD and MS-GLMB filters as discussed in Subsections~\ref{subsec:model-CPHD} and~\ref{subsec:model-MS-GLMB}. Given the posterior density, multi-target~\cite{Vo2013Labeled} or multi-trajectory~\cite{Vo2019MultiSensor, Nguyen2019GLMB} estimators can be applied to extracts target tracks. In this work, we use the sub-optimal multi-target estimators proposed in~\cite{Vo2013Labeled} to estimate tracks. The complexity of the MS-GLMB filter is $\mathcal{O}(TP^{2}(|\sum_{v=1}^{V}Z^{(v)}|))$. Hence the overall complexity of our algorithm is
 $\mathcal{O}(TP^{2}(|\sum_{v=1}^{V}Z^{(v)}|))$.
\begin{algorithm}
\caption{Robust MS-GLMB filtering algorithm}
\label{alg:robust_MSGLMB_pseudo}
\begin{algorithmic}
\Require{ modelGLMB, priorGLMB, modelCPHD, $[\text{priorCPHD}]^{(1:V)}$, $Z_{+}$}
\Ensure{$[\text{posteriorCPHD}]^{(1:V)}$, EstimatedTracks} \\
\hspace{-8pt}\textbf{\rule[0.5ex]{1\columnwidth}{0.5pt}}
\Statex
\vspace{-20pt}
\State{Compute $\tilde{\rho}^{(1)}$, $\tilde{\zeta}^{(1)}$ via Eq.\eqref{eq:GLMB-PHD} and Eq.\eqref{eq:GLMB-card} using priorGLMB.}
\For{$v=1:V$} (parallelizable)
\State{posteriorCPHD = Robust-CPHD-Recursion($\tilde{\rho}^{(1)}$, $\tilde{\zeta}^{(1)}$, modelCPHD, $\text{priorCPHD}^{(v)}$, $Z_{+})$}
\State{Compute $\bar{\kappa}^{(v)}$ via Eq.\eqref{eq:est-clutter-rate} using $\text{posteriorCPHD}^{(v)}$}
\EndFor
\State{posteriorGLMB = MS-GLMB-Recursion($\bar{\kappa}^{(1:V)}$, modelGLMB, priorGLMB, $Z_{+}$)}
\State{EstimatedTracks = Multi-Target-Estimator(posteriorGLMB)}
\end{algorithmic}
\end{algorithm}

\section{ Numerical study}\label{sec:Numerical-study}
\begin{figure}[hbt!]
    \centering \includegraphics[width=0.5\textwidth]{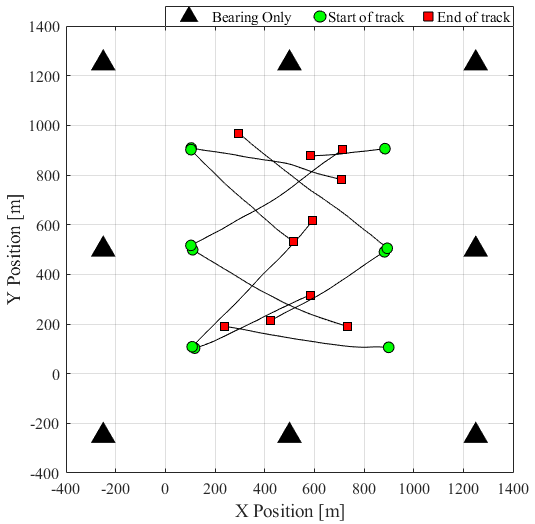} \caption{True target trajectories and sensor locations.}
    \label{fig:exp-truth}
\end{figure}

In this section, we conduct numerical studies to compare the performance of our proposed robust MS-GLMB filter to the ones of MS-GLMB (sub-optimal implementation) and iterated corrector GLMB (IC-GLMB) filters. The ground truth consists of $10$ targets in a 2-D surveillance area over a period of $100$ (s). The true target trajectories are shown in Fig.~\ref{fig:exp-truth}. A target kinematic state is represented by a 4-D vector of planar position and velocity, i.e.,  $x_{k}=[p_{x},p_{y},\dot{p}_{x},\dot{p}_{y}]^{T}$, where $T$ denotes the matrix transpose operation. The single target transition density is given by 
\begin{equation}
    f_{S+}\left(x_{+}|x,\ell \right) =\mathcal{N}\left(x_{+};Fx,Q\right),
\end{equation}
where $\mathcal{N}\left(\cdot;\bar{x},P\right)$ denotes a Gaussian distribution with mean $\bar{x}$ and covariance $P$,

$F=\left[\begin{array}{cc}
I_{2} & \Delta I_{2}\\
0_{2} & I_{2}
\end{array}\right],$ $Q=\sigma_{a}^{2}\left[\begin{array}{cc}
\frac{\Delta^{4}}{4}I_{2} & \frac{\Delta^{3}}{2}I_{2}\\
\frac{\Delta^{3}}{2}I_{2} & \Delta^{2}I_{2}
\end{array}\right]$, with $\sigma_{a}=0.15(m/s)$,  $\Delta=1(s)$ and $I_{n}$ is a $n\times n$ identity matrix. The probability of survival $p_{S}$ is set to $0.98$. The newborn targets are modeled by an LMB RFS of cardinality $6$ and the parameters are given as  $\{(r_{B+},p_{B+}^{(i)})\}_{i=1}^{6}$ where $r_{B+}=0.01$ and $p_{B+}^{(i)}=\mathcal{N}(x;m_{B+}^{(i)},Q_{B+})$ with $Q_{B+}=\textrm{diag}([10,5,10,5])$ and
 \begin{align*}
     m_{B+}^{(1)}=(100,100,0,0),&\qquad &  m_{B+}^{(2)} =(100,500,0,0),  &\qquad & m_{B+}^{(3)}=(100,900,0,0), \\
     m_{B+}^{(4)}=(900,100,0,0),&\qquad &m_{B+}^{(5)}=(900,500,0,0),&\qquad&  m_{B+}^{(6)}=(900,900,0,0).
      \end{align*}
There are eight fixed bearing-only sensors located at position $s^{(v)}=\left(s_{x}^{(v)},s_{y}^{(v)}\right),v=1,2,...,8$, 
as illustrated in Fig.~\ref{fig:exp-truth}. For sensor $v^{th}$ , the likelihood that a target $(x,\ell)$ generates
a measurement $z^{(v)}$ is given as 
\begin{equation}
g^{(v)}\left(z^{(v)}|x,\ell,s^{(v)}\right)=\mathcal{N}\left(z^{(v)};h_{\theta^{(v)}}(x,s^{(v)}),\sigma_{\theta}^{2}\right),
\end{equation}
where 
\begin{equation}
h_{\theta}\left(x,s\right)=\arctan\left(\frac{p_{x}-s_{x}}{p_{y}-s_{y}}\right),
\end{equation}
and $\sigma_{\theta}=\pi/180$ (rad). Note that MS-GLMB and robust CPHD
filters share the same single-target kinematic transition and likelihood
models.

 For a fair performance comparison, all three filters sample 3000 components during the update stage, and a maximum of 1000 components is allowed in the posterior density.
 We experimented with $100$ Monte Carlo (MC) trials. 
 The errors of the estimates are quantified via the OSPA~\cite{Schuhmacher2008AConsistent} and OSPA$^{(2)}$~\cite{Beard2020ASolution} metrics (between the estimated value and the ground truth). The cut-off and norm order of the metrics are set to $100$(m) and $1$, respectively. The window length of OSPA$^{(2)}$ metric is set to $10$(s). We perform the study on three scenarios with different simulated detection probability and clutter rate settings as given in Tab.~\ref{tab.1}. These settings are the same for all 8 sensors.
 
To show the robustness of the proposed algorithm, we keep the parameters unchanged for our filter across all scenarios. As the estimated $p_{D}$ and $\bar{\kappa}$ for all 8 sensors are similar at each scenario,  we only show the results from 1 sensor (per scenario) for demonstration purpose. 
\begin{table}[!tb]
    \centering
\begin{tabular}{|c||*{2}{c|}}\hline
\backslashbox{Scenario No.}{Parameters}
&\makebox[3em]{Detection probability $p_D$}&\makebox[3em]{Clutter rate $\bar{\kappa}^{(1,\ldots,V)}$}\\\hline\hline
1 & 0.9 & 30 \\\hline
2 & 0.5 & 5\\\hline
3 & varying between (0.5 - 0.9) & varying between (5 - 30)\\\hline

\end{tabular}
 \caption{Scenarios with different unknown parameters}
    \label{tab.1}
\end{table}
\subsection{Scenario 1} 
In this scenario, we test the performance of different filters in high clutter rate and high detection probability. For the IC-GLMB and MS-GLMB filters, we set the detection probability $p_{D}$ and clutter rate $\bar{\kappa}$ to 0.9 and 30, respectively, while these parameters are kept unknown to our filter.

\begin{figure}[hbt!]
    \centering
    \begin{subfigure}{0.49\textwidth}\centering
        \includegraphics[width=\textwidth]{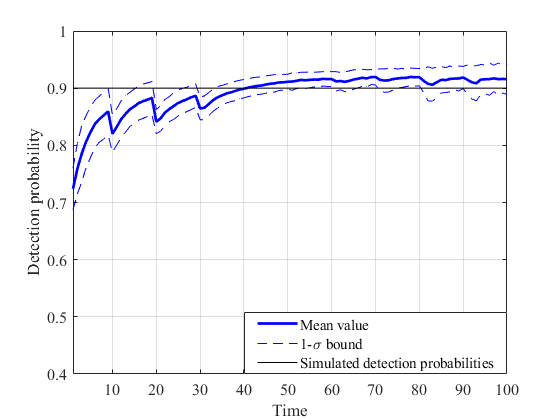} 
        \caption{Mean estimated detection probability}
    \label{est_PD_scen1}
    \end{subfigure}
        \begin{subfigure}{0.49\textwidth}\centering
        \includegraphics[width=\textwidth]{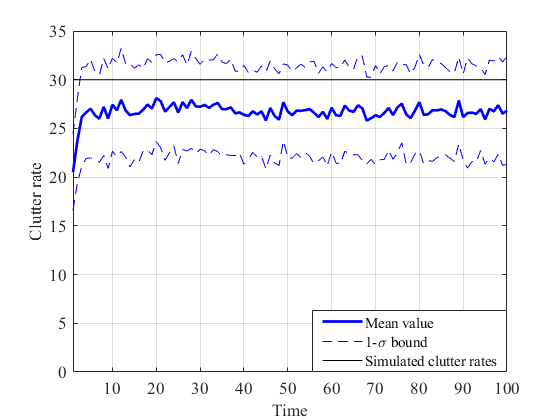}
         \caption{Mean estimated clutter rate}
               \label{est_lambda_scen1}
    \end{subfigure}
    \caption{Mean estimated detection probability and mean  clutter rate together with 0.4-$\sigma$ bound curves in the first tracking scenario.}
    \label{estimate_pD_and_lambda_scen1}
\end{figure}

Fig.~\ref{estimate_pD_and_lambda_scen1} shows the estimated detection probability and clutter rate from our filter compared to the correct values. Specifically, Fig.~\ref{est_PD_scen1} suggests that the detection probability starts at around $0.7$ (our initial settings). It approaches the correct value of $0.9$ while the estimated clutter rate is slightly below the true value.

\begin{figure}[hbt!]
    \centering
    \begin{subfigure}{0.495\textwidth} \centering \includegraphics[width=\textwidth]{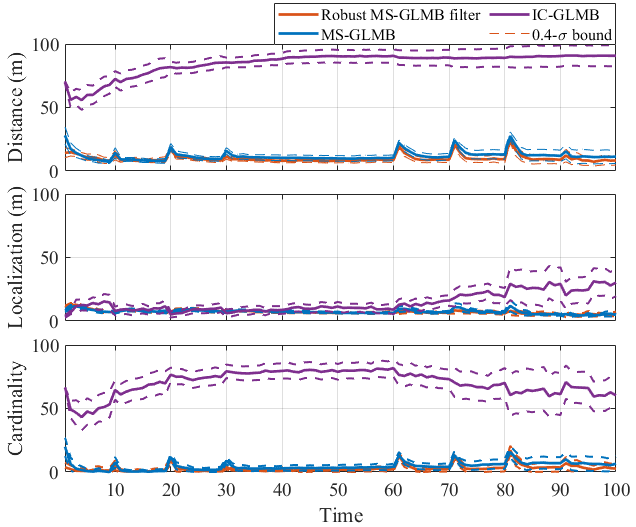}
     \caption{Mean OSPA errors}
    \label{OSPA_scen1}
    \end{subfigure} 
    \begin{subfigure}{0.49\textwidth} \centering \includegraphics[width=\textwidth]{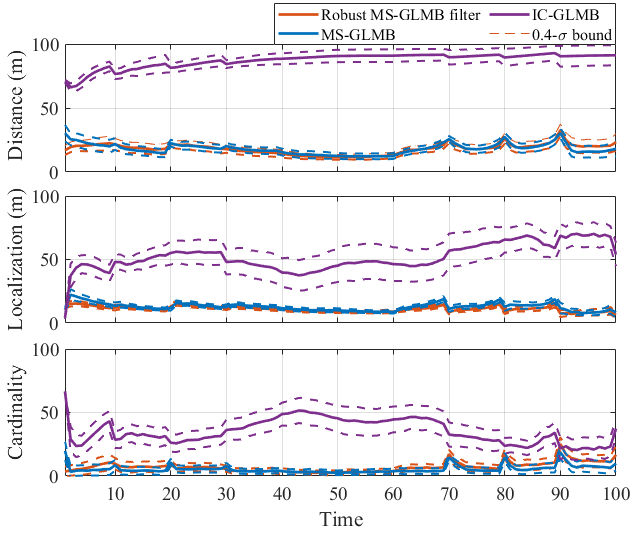}   \caption{Mean OSPA$^{(2)}$ errors} 
    \label{OSPA2_scen1}\end{subfigure}
       \caption{ Mean OSPA (left) and OSPA$^{(2)}$ (right) errors in the first tracking scenario for different filters.}
    \label{OSPA_and_OSPA2_scen1} 
\end{figure}

Regarding filtering performance, Fig.~\ref{OSPA_and_OSPA2_scen1} shows that our filter exhibits low tracking errors both in terms of localization and cardinality estimation. On the other hand, although the IC-GLMB filter gives reasonably low error in localization, it shows drastic decay in cardinality estimation quality, hence the high overall errors. Moreover, its 0.4-$\sigma$ bounds (for visualisation) over 100 MC runs show unstable tracking performance at each time step. Furthermore, the OSPA and OSPA$^{(2)}$ errors of our method and MS-GLMB filter are similar, as shown in Fig.~\ref{OSPA_scen1} and Fig.~\ref{OSPA2_scen1}. It demonstrates that our proposed method is competitive to the MS-GLMB filter. However, ours assumes no prior information on the target detection probability and clutter rate.

\begin{figure}[hbt!]
    \centering \includegraphics[width=0.5\textwidth]{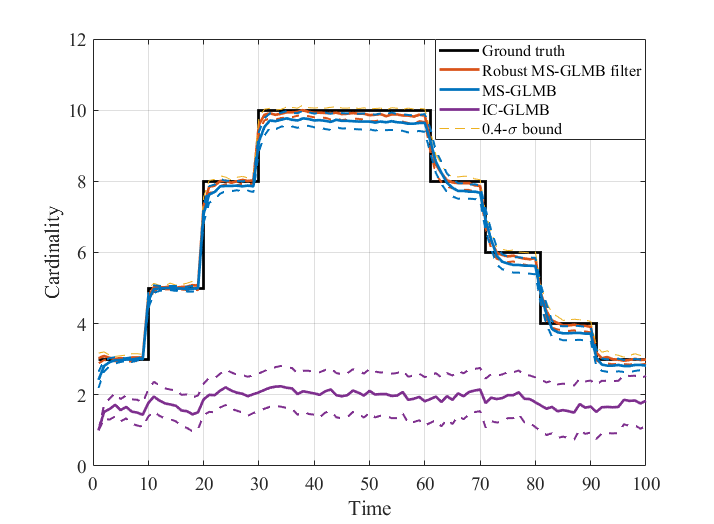} \caption{Mean estimated cardinality for different filters in scenario 1 with the ground true values.}
    \label{card_scen1}
\end{figure}
Fig.~\ref{card_scen1} shows that our filter has slightly better cardinality estimation than MS-GLMB, while IC-GLMB fails due to the high number of sensors.

\subsection{Scenario 2}

Different from Scenario 1, in this scenario, we attempt to test the tracking ability of different filters under the condition of low detection probability and low clutter rate. In this test, we set $p_D=0.5$ and $\bar{\kappa}=5$ for the IC-GLMB and MS-GLMB filters, while these parameters are estimated online in our proposed filter. 

\begin{figure}[hbt!]
    \centering
    \begin{subfigure}{0.495\textwidth}\centering
        \includegraphics[width=\textwidth]{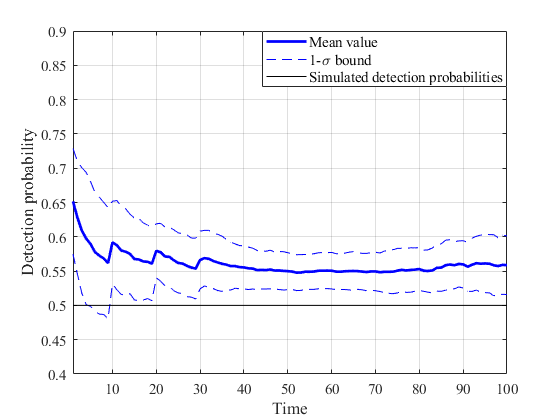} 
        \caption{Mean estimated detection probability}
    \label{est_PD_scen2}
    \end{subfigure}
        \begin{subfigure}{0.495\textwidth}\centering
        \includegraphics[width=\textwidth]{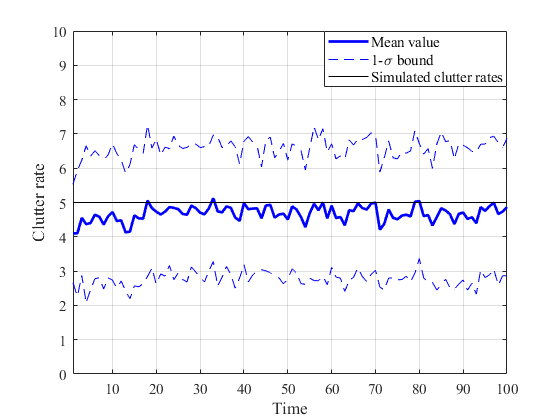}
         \caption{Mean estimated clutter rate}
               \label{est_lambda_scen2}
    \end{subfigure}
    \caption{Mean estimated detection probability and mean  clutter rate together with 0.4-$\sigma$ bound curves in the second tracking scenario.}
    \label{estimate_pD_and_lambda_scen2}
\end{figure}

Fig.~\ref{estimate_pD_and_lambda_scen2} shows that our filter correctly estimates the clutter rate of sensors (Fig.~\ref{est_lambda_scen2}), but it slightly overestimates the targets detection probability. In particular, the average detection probability starts at around $0.65$ (near our initial setting), then it gradually reduces to $0.55$ instead of the true value of $0.5$ as shown in Fig.~\ref{est_PD_scen2}.

\begin{figure}[!th]
    \centering
    \begin{subfigure}{0.495\textwidth} \centering \includegraphics[width=\textwidth]{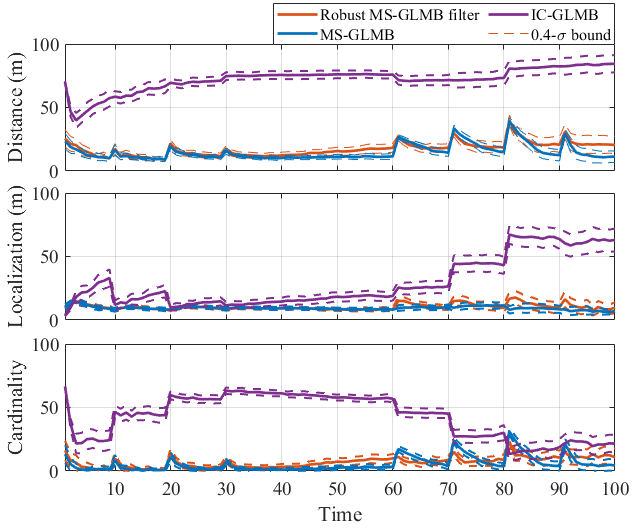}
     \caption{Mean OSPA errors}
    \label{OSPA_scen2}
    \end{subfigure} 
    \begin{subfigure}{0.495\textwidth} \centering \includegraphics[width=\textwidth]{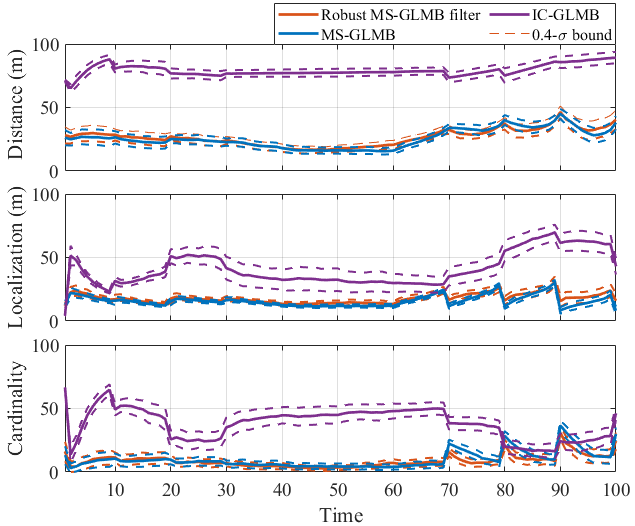}   \caption{Mean OSPA$^{(2)}$ errors} 
    \label{OSPA2_scen2}\end{subfigure}
       \caption{ Mean OSPA (left 3 plots) and OSPA$^{(2)}$ (right 3 plots) errors in the second tracking scenario for different filters.}
    \label{OSPA_and_OSPA2_scen2} 
\end{figure}

In terms of filtering performance, Fig.~\ref{OSPA_and_OSPA2_scen2} shows that IC-GLMB cannot provide reliable estimates due to a high number of sensors. On the other hand, the OSPA and OSPA$^{(2)}$ errors for our method and MS-GLMB filter are almost similar, as shown in Fig.~\ref{OSPA_scen2} and Fig.~\ref{OSPA2_scen2}. Interestingly, although we assume no prior information on the unknown background information for our filter, the OSPA$^{(2)}$ error of ours is slightly lower than that of the MS-GLMB filter. 

\begin{figure}[hbt!]
    \centering \includegraphics[width=0.5\textwidth]{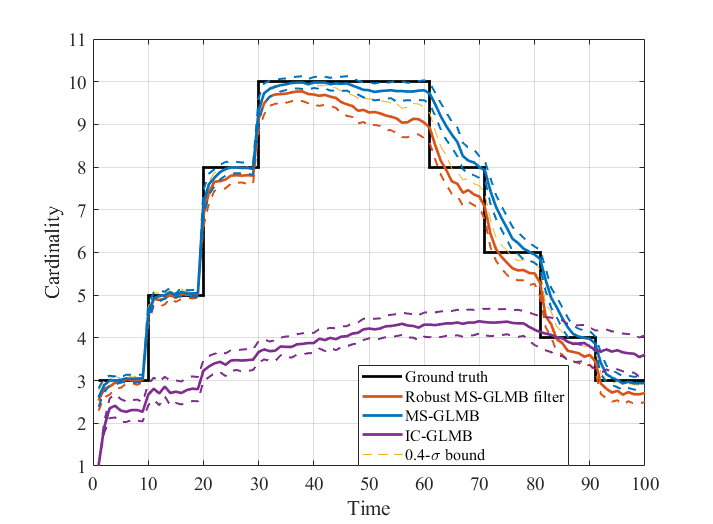} \caption{Mean estimated cardinality for different filters in scenario 2 with the ground true values.}
    \label{card_scen2}
\end{figure}

Fig.~\ref{card_scen2} shows that the MS-GLMB filter has better cardinality estimation than ours from time steps 30 to 60. From time step 60 onward, the MS-GLMB filter tends to provide overestimation while ours slightly underestimates the number of targets. IC-GLMB cannot reliably track targets in this scenario.

\subsection{Scenario 3}

To test the filters in time-varying detection profile and clutter rate condition, we create a scenario where the detection probability $p_D$ and clutter rate $\lambda_c$ values increase from low to high during tracking period. Specifically, for the first 50 time steps, we simulate the condition with low detection probability and low clutter rate (parameters as in the scenario 2). For the remaining tracking period, we simulate tracking condition with high detection probability and high clutter rate (parameters as in the scenario 1). The average detection probability and clutter rate for IC-GLMB and MS-GLMB are set to the mean values of the variation ranges, i.e. 0.7 and 17.5 for detection probability and clutter rate, respectively.
\begin{figure}[hbt!]
    \centering
    \begin{subfigure}{0.495\textwidth}\centering
        \includegraphics[width=\textwidth]{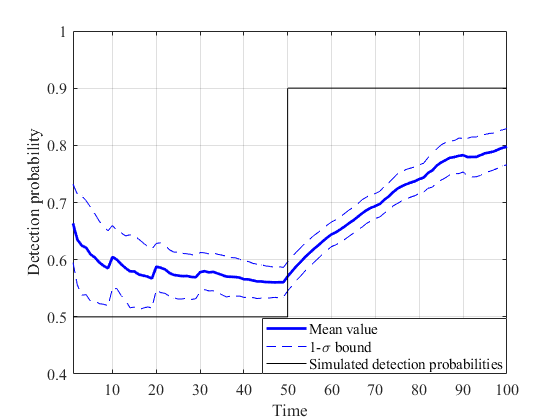} 
        \caption{Mean estimated detection probability}
    \label{est_PD_scen3}
    \end{subfigure}
        \begin{subfigure}{0.495\textwidth}\centering
        \includegraphics[width=\textwidth]{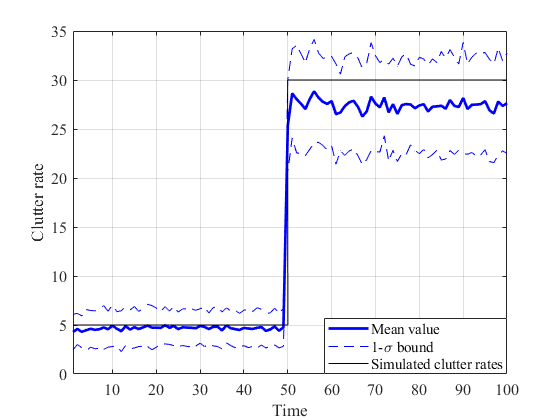}
         \caption{Mean estimated clutter rate}
               \label{est_lambda_scen3}
    \end{subfigure}
    \caption{Mean estimated detection probability and mean  clutter rate together with 0.4-$\sigma$ bound curves in the third tracking scenario.}
    \label{estimate_pD_and_lambda_scen3}
\end{figure}
\begin{figure}[hbt!]
    \centering
    \begin{subfigure}{0.495\textwidth} \centering \includegraphics[width=\textwidth]{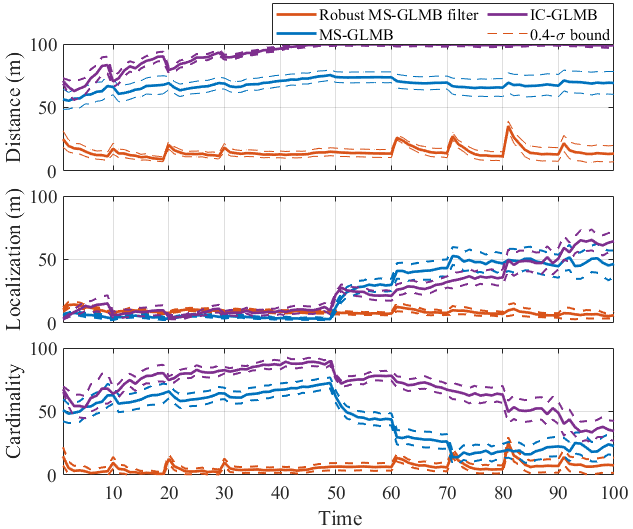}
     \caption{Mean OSPA errors}
    \label{OSPA_scen3}
    \end{subfigure} 
    \begin{subfigure}{0.495\textwidth} \centering \includegraphics[width=\textwidth]{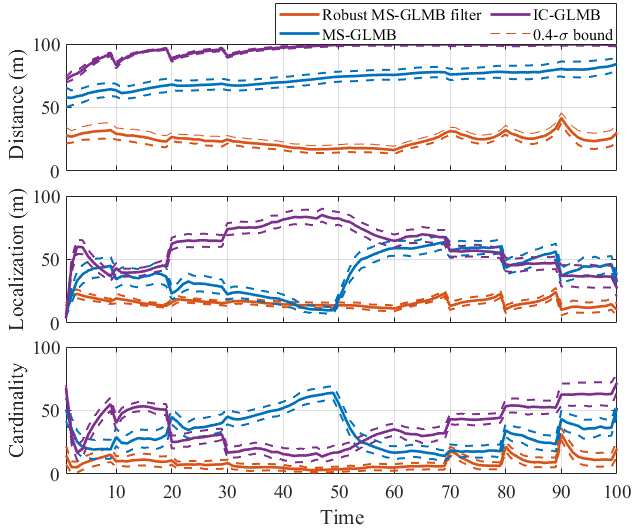}   \caption{Mean OSPA$^{(2)}$ errors} 
    \label{OSPA2_scen3}\end{subfigure}
       \caption{ Mean OSPA (left 3 plots) and OSPA$^{(2)}$ (right 3 plots) errors in the third tracking scenario for different filters.}
    \label{OSPA_and_OSPA2_scen3} 
\end{figure}
  
Fig.~\ref{estimate_pD_and_lambda_scen3} presents the estimated detection probability and clutter rate from our proposed filter. It demonstrates that our filter correctly captures the variations of background condition. It is also observed that the sensitivity of our filter to the change in clutter rate is higher than that of the one in detection probability.

\begin{figure}[hbt!]
    \centering \includegraphics[width=0.5\textwidth]{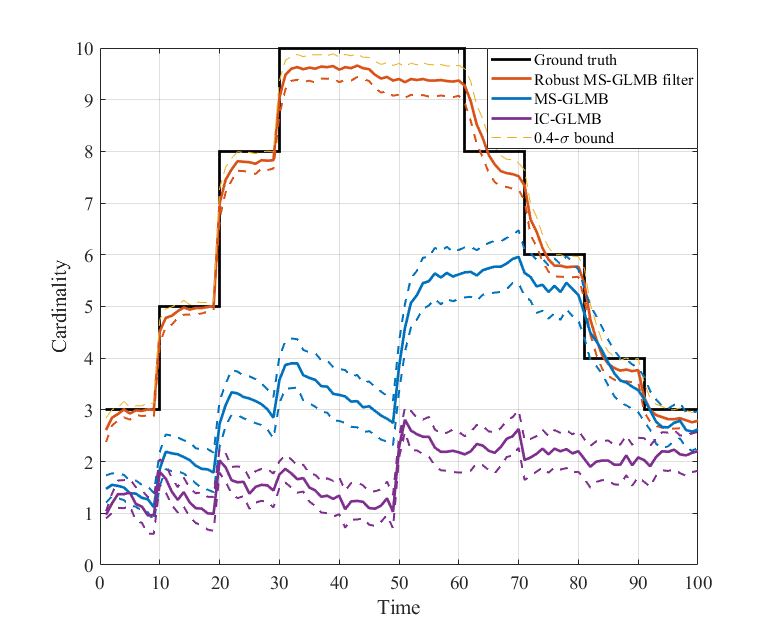} \caption{Mean estimated cardinality for different filters in scenario 3 with the ground true values.}
    \label{card_scen3}
\end{figure}

Considering the filtering performance in terms of OSPA and OSPA$^{(2)}$ errors, the results in Fig.~\ref{OSPA_and_OSPA2_scen3} shows that although MS-GLMB filter has better performance than the IC-GLMB, both of these filters fail to track targets in this scenario. Fig.~\ref{OSPA_and_OSPA2_scen3} also indicates that our filter tracks all targets with high level of accuracy. This is due to the capability of our filter in adapting to the changes of background condition.

\begin{table}[!tb]
    \centering
\begin{tabular}{|c||*{3}{c|}}\hline
\backslashbox{Scenario No.}{Filters}&{IC-GLMB}&{MS-GLMB}&{Robust MS-GLMB}\\\hline\hline
1 & 94.44($\pm$ 7.74) & \textbf{28.81}($\pm$ 10.45) & 40.51 ($\pm$ 10.55)\\\hline
2 & 93.24 ($\pm$ 3.61) & 47.63 ($\pm$ 9.83) & \textbf{46.59} ($\pm$ 12.36)\\\hline
3 & 99.38 ($\pm$ 1.22) & 84.31 ($\pm$ 3.91) & \textbf{30.16} ($\pm$ 10.62) \\\hline
\end{tabular}
 \caption{ OSPA$^{(2)}$ errors ($\pm$ 1-$\sigma$ bound) (m) (over 100 MC runs evaluated over the entire tracking period (window length of 100s) of different filters in all scenarios.} 
    \label{tab.2}
\end{table}

Fig.~\ref{card_scen3} shows that IC-GLMB and MS-GLMB filters incorrectly estimate the targets set cardinality in this scenario. On the other hand, our method demonstrates reliable estimation results, although it tends to slightly underestimate the number of targets from time step 30 onward.

For overall comparison, Table~\ref{tab.2} shows the mean OSPA$^{(2)}$ errors (over $100$ MC runs) evaluated over the entire tracking period. The results show that the IC-GLMB filter has the worst performance in all three scenarios. On the other hand, the performance of our proposed robust filter is worse than of the MS-GLMB filter in the first scenario. However, it is competitive to the MS-GLMB filter in the second scenarios (which is supplied with correct background parameters). For the last scenario, our proposed filter has  the best performance compared to the ones of others given its capability to adapt to the changes in tracking environment.

Although the targets are accurately tracked, we observe the  mean detection probability is overestimated while the clutter rate is underestimated.  One hypothesis is that in some GLMB components, clutter measurements might be associated with  true miss-detected tracks (those components might have small weights). In scenario 1, since the number of clutter measurement is high, the underestimation of clutter rate is more severe. On the other hand, due to the characteristics of the beta distribution\footnote{In low detection probability scenarios, a detection event will increase the mean detection probability more significant than in high detection probability scenarios.} and the high detection probability of the scenario, the underestimation of detection probability is not noticeable. Similar explanation can also be established for scenario 2, but since the clutter rate is low, the underestimation is not noticeable. However, due to the low detection probability of the scenario, the detection probability overestimation is more severe.

\section{Conclusions}\label{sec:Conclusions} 

This paper has proposed a robust multi-sensor multi-target tracking algorithm based on the MS-GLMB filter for multi-target tracking and the low-cost robust CPHD filters for clutter rate estimation. Given this formulation, the proposed algorithm efficiently estimates the target trajectories and background information online. The experimental results show that our method provides reliable estimates in different tracking conditions with multiple bearing-only sensors, whereas the iterated corrector approach fails completely due to the high number of sensors. Notably, our filter has a similar performance to the MS-GLMB filter supplied with the correct background parameters in constant background condition. In scenario with fluctuating backgrounds, our filter outperforms the MS-GLMB filter because it can learn and adapt to the changes in clutter rate and target detection profile.

The proposed filter currently assumes prior knowledge of the newborn targets distribution. In practice, this information is not available in many applications. While the measurement adaptive birth model~\cite{Reuter2014TheLabeled} can be readily extended to a multi-sensor setting, it is expensive to combine measurements from different sensors to form the birth density. Hence, an efficient method for births modelling will be considered for a more adaptive multi-sensor multi-target tracking algorithm for future work.
\section*{Acknowledgment}

The authors acknowledge the supervision of their supervisors and mentors for this work.

 \bibliographystyle{IEEEtran}
 \bibliography{IEEEabrv,refbib2021}

\end{document}